\documentclass[aps,twocolumn,nofootinbib]{revtex4}

\usepackage{graphicx}
\usepackage{epic}
\usepackage{eepic}
\usepackage{latexsym}

\newcommand{\eq}[1]{(\ref{#1})}
\newcommand{\be}{\begin{equation}}
\newcommand{\ee}{\end{equation}}
\newcommand{\bea}{\begin{eqnarray}}
\newcommand{\eea}{\end{eqnarray}}

\newcommand{\vs}[1]{\vspace{#1 mm}}
\newcommand{\hs}[1]{\hspace{#1 mm}}

\def\a{\alpha}
\def\b{\beta}

\def\d{\delta}

\def\f{\phi}
\def\fr{\frac}

\def\vf{\varphi}

\def\l{\lambda}

\def\m{\mu}
\def\n{\nu}

\def\s{\sigma}

\def\th{\theta}

\def\O{\Omega}

\def\del{\partial}

\let\bm=\bibitem
\def\nn{\nonumber}
\def\vf{\varphi}

\begin{document}

\title{S-duality in String Gas Cosmology}

\author{Savas Arapoglu$^{1}$}
\email[]{arapoglu@gursey.gov.tr}
\author{Ata Karakci$^{2}$}
\email[]{karakcia@boun.edu.tr}
\author{Ali Kaya$^{1,2}$\vs{3}}
\email[]{ali.kaya@boun.edu.tr}
\affiliation{$^{1}$Feza G\"{u}rsey Institute,\\
Emek Mah. No:68, \c{C}engelk\"{o}y, \.Istanbul, Turkey\vs{3}\\
$^{2}$Bo\~{g}azi\c{c}i University, Department of Physics, \\ 34342,
Bebek, \.Istanbul, Turkey\vs{3}}

\date{\today}

\begin{abstract}

We consider a toy cosmological model in string theory involving the
winding and momentum modes of  $(m,n)$ strings, i.e. bound states of
$m$ fundamental and $n$ D-strings. The model is invariant under
S-duality provided that $m$ and $n$ are interchanged. The dilaton is
naturally stabilized due to S-duality invariance, which offers a new
mechanism of moduli fixing in string gas cosmology. Using a tachyon
field rolling down to its ground state, we also point out a possible
way of realizing a cosmological phase  with decreasing
Hubble radius and constant dilaton.

\end{abstract}

\maketitle

\section{Introduction}

String gas cosmology (SGC) \cite{bv} is a natural way of applying
string  theory to cosmology (for review see  \cite{rev1,rev2}). In addition to usual field theoretic
excitations, in SGC one also considers new degrees of freedom of string theory
and it turns out that the winding modes, which exist in the presence of compact directions, play an important cosmological role.
The stringy T-duality invariance is the key symmetry which may
resolve the initial big-bang singularity of standard cosmology.
Moreover, the naive dependence of the tree level annihilation cross
section of winding strings on the number of dimensions might explain
the hierarchy between the observed three and internal directions
\cite{bv} (however it seems there are some  caveats in that argument
\cite{st0,st1,st2,cam} and a more complete understanding of string
thermodynamics in an expanding universe is needed). In late times,
SGC offers a natural mechanism to stabilize extra dimensions as a
result of a balance between negative winding and positive momentum
pressures \cite{ss1}. Assuming that the cosmology is dominated by
states which become massless at the special self-dual
compactification radius, one can naturally avoid exceeding the
phenomenological bounds on the energy density of the universe
\cite{sm0,sm1,sm2}. A string gas with fluxes is also capable of
fixing the flux and the shape moduli \cite{sf}.

SGC was developed in \cite{b1,b2,b3} to include higher dimensional
$p$-branes that exist in string/M theory.  Although, these are expected
to annihilate in early times according to Brandenberger-Vafa
mechanism \cite{bv}, higher dimensional branes are needed to
stabilize topologically non-toroidal cycles since winding string modes do not
exist in the spectrum when the first homology class vanishes.
Indeed, $p$-branes are capable of fixing not only the volume modulus of an
internal Ricci flat manifold (radion) \cite{bs1,bs2} (see also
\cite{t1,t2,t3}) but also the shape modulus of an internal torus
they wrapped by the stress they generate \cite{bshape}.

Despite the fact that the late time stabilization of extra
dimensions is natural in string/brane  gas cosmology, dilaton
stabilization (or stabilization of dilaton and radion
simultaneously) is problematic \cite{stab1,stab2} (see however \cite{stab3} and \cite{stab2} for a solution by superpotentials).  As T-duality
plays a key role for the stabilization of  extra dimensions,
S-duality, which basically acts on  dilaton, may assist in fixing
dilaton. Indeed it was shown in \cite{ss} that compactifying the eleven dimensional supergravity on a circle  the  cosmological membrane production in the enhanced symmetry point can force the radius (and thus dilaton) towards the self S-dual point\footnote{$(m,n)$ strings that we consider in this paper  also arise in the  membrane  compactification studied in \cite{ss}.}. 

Another motivation to consider S-duality in SGC comes from the
recently proposed stringy mechanism to generate scale free
cosmological perturbations seeded by thermal fluctuations in a
strongly coupled  Hagedorn phase \cite{pert1,pert2,pert3,pert4}.
According to \cite{pert1,pert2,pert3,pert4}, spacetime should be static in
this meta-stable phase with infinite Hubble radius and it should be
followed by the so called Hagedorn phase which has decreasing Hubble
radius. During the Hagedorn phase cosmological perturbations exit
the Hubble radius and freeze out. As criticized in \cite{crt}, for
this mechanism to really work out, the dilaton should be constant so
that the string and Einstein frame Hubble parameters coincide. A
possible way out of this criticism is to assume that the strongly
coupled Hagedorn phase corresponds to a self S-dual state and thus
dilaton should be equal to zero \cite{pert4}.

In the absence of  a special symmetry  that implies non-renormalization
(like supersymmetry), it seems
difficult to give an effective field theory description of the
strongly coupled Hagedorn phase since perturbative field equations
cannot be trusted. However, it should be possible to model the
Hagedorn phase even if one enters into the strong coupling regime
above the zero dilaton, since in that case one can apply an
S-duality map to yield a weak coupling description.

Alternatively, one can also consider the possibility of having constant dilaton also in the Hagedorn phase. In that case, the equations of dilaton gravity reduce to that of usual Einstein's equations. Although it is not directly related to the main interest of this paper, namely S-duality, we also point out that using a tachyon field rolling down to its ground state it is possible to obtain a cosmological evolution with decreasing Hubble radius which can be matched to the static strongly coupled Hagedorn phase.

\section{S-duality}

Let us start with the low energy effective action in 10-dimensions
\be
S_{10}=\fr{1}{l_s^8}\int \sqrt{-g} e^{-2\f}\left[R+4 (\nabla\f)^2\right].\label{s1}
\ee
It is well known that \eq{s1} is invariant under S-duality transformation
\bea
&&\f\to -\f\, ,\nn\\
&&g_{\m\n}\to e^{-\f}g_{\m\n}\label{s}.
\eea
Since the string coupling constant $g_s$ is given by
$g_s=e^\f$, this map corresponds to a strong-weak coupling
duality. Assuming a metric of the form,
\be
ds^2=-e^{2A}dt^2+\sum_i e^{2B_i} d\th_i^2, \label{3}
\ee
where the functions $A,B_i$ depend only on time $t$,
the action \eq{s1} reduces to
\be
S_{10}=l_s \int dt \,\, e^{-\varphi-A}\left[\sum_{i} \dot{B}_i^2 -\dot{\varphi}^2\right],\label{s14}
\ee
where the shifted dilaton is defined as
\be
\varphi=2\f-\sum_i B_i.
\ee
Here we take the dimensionful coordinates $\th_i$ to have length one in string units.

To add stringy matter to \eq{s1}, the corresponding free energy $F$ should be coupled minimally to the string metric in the form of an effective Lagrangian \cite{tv}. Differing from \cite{tv}, we assume that the free energy also depends on
the string coupling constant, hence dilaton (or shifted dilaton), and the total action becomes
\be
S=S_{10}-\int dt\, e^A \,F(B_i,\vf,\b e^A).\label{tf}
\ee
The field equations following from this action in the proper time $A=0$ (and in string units $l_s=1$)  read
\bea
&&\dot{\vf}^2-\sum_i \dot{B}_i^2=e^\vf E,\nn\\
&&\ddot{B}_i-\dot{\vf}\dot{B}_i=\fr{1}{2}e^\vf P_i, \label{f1}\\
&& 2 \ddot{\vf}-\dot{\vf}^2-\sum_i \dot{B}_i^2=-e^\vf P_\vf  ,\nn
\eea
where
\be
E=F+\b\fr{\del F}{\del \b},\hs{5}
P_i=-\fr{\del F}{\del B_i},\hs{5}
P_\vf=-\fr{\del F}{\del \vf}.
\ee
As a direct consequence of field equations, the coupled
sources should obey the energy-momentum  conservation equation
\be
\dot{E}+\sum_i \dot{B}_i P_i +\dot{\vf} P_\vf=0.\label{con}
\ee
Here, we see that the shifted dilaton looks like an extra dimension,
which is not surprising due to  the 11-dimensional origin of dilaton.

The conservation equation \eq{con} implies that the entropy
\be
S\equiv\b^2\fr{\del F}{\del \b}
\ee
is a constant of motion, hence the evolution is adiabatic.
The temperature will adjust itself to yield constant entropy and
one can write $\b=\b(B_i,\vf)$. Therefore in the adiabatic approximation energy $E$
can be viewed as a function of $B_i$ and $\vf$. In that 
case the pressures can be calculated
as\footnote{To verify this note that when $\b=\b(B_i,\vf)$ and $S=S_0$,
$E=\hat{F}+S_0/\b$ where $\hat{F}(B_i,\vf)= F(B_i,\vf,\b(B_i,\vf))$.
Then $\fr{\del E}{\del B_i}=\fr{\del \hat{F}}{\del B_i}-
\fr{\del \b}{\del B_i}\fr{S_0}{\b^2}=\fr{\del F}{\del B_i}
=-P_i$. The same manipulations for $\vf$ gives \eq{pe}.}
\be
P_i=-\fr{\del E}{\del B_i}, \hs{5}P_\vf=-\fr{\del E}{\del \vf},\label{pe}
\ee
which means that  the total action becomes equivalent to 
\be
S=S_{10}-\int dt\, e^A \,E(B_i,\vf).
\ee
The action will be T-duality invariant provided
\be
E(B_i,\vf)=E(-B_i,\vf).
\ee
On the other hand, from \eq{s}, S-duality invariance requires
\be
E(B_i,\f)=e^{\f/2}E(B_i-\f/2,-\f).
\ee
Note that in this equation we treat $E$ as a function of $\f$, not the shifted dilaton $\vf$.

\section{Stabilization of dilaton}

In type IIB string theory, an S-duality transformation \eq{s}
converts a fundamental string into a D-string. Actually, the action
of S-duality in type IIB theory is more general than the mere
strong-weak coupling map \eq{s}. It can be represented as an
$SL(2,Z)$ transformation and by acting on the fundamental string one
can obtain an $SL(2,Z)$ multiplet of strings  
characterized by two integers $m$ and $n$ \cite{schw}. An  $(m,n)$ string can
be viewed as the bound state of $m$ fundamental and $n$ D-strings
\cite{wit}. The transformation \eq{s} simply interchanges $m$ and
$n$. In string frame the tension of the $(m,n)$ string is given
by \cite{schw}
\be
T=T_s \sqrt{m^2+\fr{n^2}{g_s^2}},\label{ten}
\ee
where $g_s$ is the string coupling constant and $T_s$ is the tension of the fundamental $(1,0)$ string determining the string length $l_s$. In seeking how
S-duality acts in the context of string gas cosmology,
it is clear that one should study a model involving
$(m,n)$ strings.

The energy of an $(m,n)$ string winding a compact direction is
proportional to its tension \eq{ten} times the radius of the circle.
On the other hand, the energy of a momentum mode (either
corresponding to a small vibration of a winding string or to an
unwound string circling around that direction) is  inversely
proportional to the compactification radius. Note that the energy of
the momentum mode should not depend on the tension \eq{ten}. It
should be measured in units of the corresponding angular coordinate
parametrizing the circle and thus from \eq{s1}  and \eq{s14} in
terms of the string length $l_s$.

As a result, the string frame matter action for the winding and momentum modes of $(m,n)$ strings should take the form
\bea
S=-\sum_i \int dt \left( E_w\sqrt{m^2+n^2 e^{-2\f}}\,\, e^{A+B_i}+E_m\, e^{A-B_i}\right)\nn,
\eea
where the sum should go over {\it compact directions}, and $E_w$ and $E_m$ are  constants characterizing winding and momentum energies, respectively. It is easy to see that the matter action is S-duality invariant with $m\leftrightarrow n$.

The S-duality transformation \eq{s} is more  manifest in Einstein frame  defined by
\be
g_{\m\n}^{(E)}=e^{-\f/2}g_{\m\n}.
\ee
Writing the Einstein frame metric as (note that we are using the same letters for the metric components in Einstein frame, which should not be confused with \eq{3})
\be
ds_E^2=-e^{2A}dt^2+\sum_i e^{2B_i} d\th_i^2,
\ee
the total action becomes
\bea
&&S_E=l_s\int  dt \,\, e^{-A+\sum_j B_j}\left[\sum_{i} \dot{B}_i^2 -(\sum_j \dot{B}_j)^2+\fr12 \dot{\f}^2
\right]\nn\\
&& -\sum_i \int dt \left( E_w\sqrt{m^2e^{\f}+n^2 e^{-\f}}\,\, e^{A+B_i}+E_m\, e^{A-B_i}\right)\nn.
\eea
S-duality transformation in the Einstein frame is simply given by 
\be
\f\to-\f, \label{sde}
\ee
and the invariance of the action $S_E$ is obvious provided $m\leftrightarrow n$. Note that S-duality leads to a potential\footnote{Actually $V(\f)$ is not an honest potential since the action is not of the form $\int  V\sqrt{-g}$. However, it still implies a non-trivial dilaton dependence of non-kinetic energy.}  for dilaton $V(\f)=E_w\sqrt{m^2e^{\f}+n^2 e^{-\f}}$ which has a global minimum (see figure). 

\begin{figure}
\centerline{
\includegraphics[width=7.0cm]{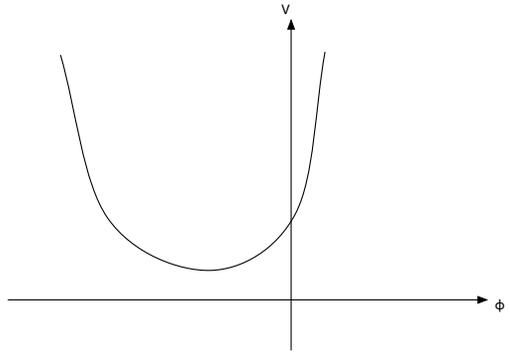}}
\caption{The dilaton potential for $m>n$.}
\end{figure}

In the proper time, the dilaton equation that follows from the above action reads (we set $l_s=1$) 
\be
\ddot{\f}+\dot{k} \dot{\f}=-\fr{1}{2}E_w \sum_i e^{B_i-k}\fr{m^2 e^{\f}-n^2 e^{-\f}}{\sqrt{m^2e^{\f}+
n^2 e^{-\f}}}, \label{de}
\ee
where
\be
k=\sum_i B_i .
\ee
From \eq{de} one sees that the constant dilaton
\be
\f=\ln (n/m)\label{f0}
\ee
is a solution. Moreover, \eq{f0} is also consistent with other field equations. Note that it is possible to stay in the weak coupling by choosing $m\gg n$.

To see that this is a stable point for dilaton and hence it is fixed, we first point out that $\dot{k}$ never vanishes, which follows from the variation of the action with respect to $A$.  To preserve adiabatic approximation we impose $\dot{k}>0$ and thus $\dot{k}\dot{\f}$  term in \eq{de} acts like a frictional force that dumps the motion.
Moreover, from the right hand side we see that when $\f>\ln (n/m)$   there is a negative and when $\f<\ln (n/m)$ there is a positive acceleration that force the equilibrium value. Indeed, looking for a small perturbation $\d\f$ around \eq{f0} we find that it obeys
\be
\ddot{\d\f}+\dot{k} \dot{\d\f}=-\left(E_w\,\sqrt{\fr{mn}{2}} \sum_i e^{B_i-k}\right) \d\f,
\ee
which is similar to a dumped oscilator with time dependent friction and frequency.

To quantify these arguments and to see whether it is possible to stabilize both dilaton and radion simultanously let us consider $(1+d+p)$ dimensional split of the  spacetime with the following metric
\be
ds_E^2=-dt^2+e^{2B}ds_d^2+e^{2C}ds_p^2.
\ee
Here, $ds_d^2$ and  $ds_p^2$ are flat metrics on $R^d$ and $T^p$, playing the roles of the observed and internal dimensions. From the Einstein frame action it is possible to obtain the following field equations
\bea
&&\ddot{\f}+\dot{k} \dot{\f}=-\fr{p}{2}E_w e^{C-k}\fr{m^2 e^{\f}-n^2 e^{-\f}}{\sqrt{m^2e^{\f}+n^2 e^{-\f}}},\nn\\
&&\ddot{C}+\dot{C} \dot{k}=F_m-\fr{(d-5)}{4} F_w,\label{fe}\\
&&\ddot{B}+\dot{B} \dot{k}=\fr{p}{4} F_w,\nn \eea
together with the
initial constraint
\be
\dot{k}^2=d\dot{B}^2+p\dot{C}^2+\fr{1}{2}\dot{\f}^2+2p F_w+2p F_m,
\ee where the functions $F_w$ and $F_m$ are given by \bea
&&F_w=\fr{E_w}{2}\sqrt{m^2e^{\f}+n^2 e^{-\f}}e^{-k+C},\nn\\
&&F_m=\fr{E_m}{2}e^{-k-C}.
\eea
Note that $k=d B+p C$ and $d+p=9$.

For $d>5$, there is a special solution with constant dilaton and radion given by
\be
\f=\ln (n/m),\hs{5} B=\fr{2}{d}\ln (\a t),\hs{5}C=C_0,
\ee
where
\bea
&&e^{2C_0}=\fr{8E_m}{(d-1)E_w\sqrt{2mn}},\nn\\
&&\a^2=\fr{dpE_w\sqrt{2mn}}{16}\,e^{(1-p)C_0}.
\eea
To verify stabilization let us consider small perturbations around this background which obey
\bea
&&\ddot{\d\f}+\fr{2}{t}\dot{\d\f}+\fr{8}{d}\,\fr{\d \f}{t^2}=0,\nn\\
&&\ddot{\d C}+\fr{2}{t}\dot{\d C}+\fr{4(d-5)}{dp}\,\fr{\d C}{t^2}=0,
\label{fcpert}\\
&&\ddot{\d B}+\fr{4}{t}\dot{\d B}+\fr{2}{t^2}\d B+\fr{2p}{d}\fr{\dot{\d C}}{t}+\fr{(2p-2)}{d}\,\fr{\d C}{t^2}=0.\nn
\eea
We see that $\d \f$ and $\d C$ are nicely decoupled. The corresponding linearly independent solutions for them can be found as 
\be
\fr{\cos\left[\ln\left(\fr{t\sqrt{4\l-1}}{2}\right)\right]}{\sqrt{t}},\hs{5}
\fr{\sin\left[\ln\left(\fr{t\sqrt{4\l-1}}{2}\right)\right]}{\sqrt{t}},
\ee
where $\l=8/d$ for $\d \f$ and $\l=4(d-5)/(dp)$ for $\d C$, and in each case  $\l>1/4$.
Thus both $\d\f$ and $\d C$ fall for large $t$ as $1/\sqrt{t}$,
which proves that they are stabilized. Similarly, it can be seen from \eq{fcpert} that the perturbation $\d B$ also goes like $1/\sqrt{t}$ as $t\to\infty$.

One may think that  a cosmology dominated by $(m,n)$ strings of one type  is not realistic. Indeed, $(m,n)$ strings have a larger tension compared to, say, $(1,0)$ or $(0,1)$ strings, which become lightest strings in Einstein frame at weak and strong couplings, respectively. However, assuming  that the numbers of fundamental and D-strings do not change and, for the moment, they are equal to each other, then it is energetically more favorable to form  $(1,1)$ strings, since the tension of a $(1,1)$ string is less than the sum of the tensions of $(1,0)$ and $(0,1)$ strings. Similarly, when the ratio of the numbers is $m/n$, the cosmology will be dominated by $(m,n)$ strings. 

The restriction $d>5$ on the number of observed dimensions in the above toy model can easily be circumvented by including higher dimensional S-duality related branes. For instance, in type IIB string theory one can consider a bound state of $m$ Dirichlet and $n$ solitonic fivebranes which has the following tension in the Einstein frame \cite{wit}
\be
T=T_5 \sqrt{m^2 g_s+\fr{n^2}{g_s}}.
\ee
As for strings, the duality map \eq{sde} interchanges $m$ and $n$. One can consider  $(1+3+5+1)$ dimensional split of the spacetime where $(m,n)$ fivebranes and strings are wrapping over the  five and the one dimensional subspaces of the  six dimensional toroidal internal space, respectively. It  can be verified that in this S-duality invariant model both radion and dilaton  are fixed in a three dimensional observed space, where again $\f=\ln(n/m)$.

The above results indicate that S-duality can  play a crucial role in SGC, especially  in the late time stabilization of dilaton. For early time applications, it seems a detailed information about the free energy of the string gas, particularly its  dependence on the string coupling constant, is needed.

\section{Discussion}

As a final comment we would like to point out a possible way of getting a decreasing Hubble radius in string theory. This should mimic the cosmological evolution in the Hagedorn phase \cite{pert4}, following the static strongly coupled Hagedorn phase.
We assume that the dilaton and internal dimensions are already stabilized leaving a four dimensional Einstein theory. A decreasing Hubble radius necessarily implies accelerated expansion, but the opposite is not correct. Therefore, it seems harder to obtain former compared to the acceleration. Indeed, if the spatial sections of the universe are flat, one can only get a decreasing Hubble radius by violating the null energy condition, i.e. $\rho+P<0$.

One method of obtaining acceleration in string theory is via a tachyon  field $T$ rolling down to its ground state \cite{gib}. The tachyon potential $V(T)$ has a positive maximum $V_0$ at $T=0$ and a minimum at $T=T_0$ with $V(T_0)=0$. The Lagrangian is given by
\be
{\cal L}=-V(T)\sqrt{1-\dot{T}^2}.
 \ee
Considering the evolution of tachyon field from  $T=0$ (with a small initial velocity) towards the minimum at $T=T_0$, one sees that $\dot{T}$ increases in time. From the corresponding energy momentum tensor it follows  that for  $\dot{T}^2<2/3$, $\rho+3P<0$ which implies acceleration \cite{gib}. However, in this interval $\rho+P>0$ which gives an increasing Hubble radius. Thus the tachyon field alone cannot yield a decreasing Hubble radius.

Actually, there is another problem here which is also crucial for our discussion. From field equations one can see that during this accelerating phase the expansion speed can never vanish. Therefore, it is not possible to match this evolution to the static strongly coupled Hagedorn phase. To cure this difficulty one should essentially try to slow down the expansion speed. To that effect, the easiest madification  is to assume that the space is positively curved rather than being flat. Taking the spacetime metric as
\be
ds^2=-e^{2A}dt^2+e^{2B}d\O_3^2,
\ee
where $d\O_3^2$ is the unit round metric on the three sphere, and coupling the tachyon field with the canonical potential $V(T)$, one obtains the action
\bea
S_4&=&l_p\int dt\, e^{-A+3B}  (-6\dot{B}^2)+6\s e^{A+B}\nn\\
&-&\int dt \, e^{A+3B}\, V(T)\,\sqrt{1-e^{-2A}\dot{T}^2},
\eea
where  $\s=1$ and that term is related to the curvature of the sphere. Varying with respect to $A,B$ and $T$ one finds (we set $l_p=1$) 
\bea
&&\dot{B}^2=-\s e^{-2B}+\fr{V}{6\sqrt{1-\dot{T}^2}},\label{tac1}\\
&&\ddot{B}=\s e^{-2B}-\fr{V\dot{T}^2}{4\sqrt{1-\dot{T}^2}},\label{tac2}\\
&&\fr{d}{dt}\left(\fr{V}{\sqrt{1-\dot{T}^2}}\right)=-3\dot{B}
\fr{V\dot{T}}{\sqrt{1-\dot{T}^2}}.\label{tac3}
\eea
Note that, although the curvature of the sphere ($\s$-term) contributes negatively
to the expansion speed $\dot{B}$, it  increases  $\ddot{B}$, which is precisely what we were looking for.

Since we would like to match the background to the static strongly coupled Hagedorn phase, we assume that initially at $t=0$, $\dot{B}(0)=0$. For the tachyon field we impose $T(0)=0$ and take a small positive initial velocity, i.e $0<\dot{T}(0)\ll 1$ (assuming $\dot{T}(0)=0$ gives the Einstein static universe which is unstable). Then from \eq{tac1} and \eq{tac2} one sees that $\ddot{B}(0)>0$. Recalling that the Hubble radius $R_H$ is given by $R_H\sim 1/\dot{B}$, it decreases initially.

To understand the succeeding  evolution we note that since $\ddot{B}+\dot{B}^2\sim (1-3\dot{T}^2/2)$ the acceleration stops when $\dot{T}^2=2/3$. Thus,  $\ddot{B}$ should turn to negative at some earlier time and subsequently the Hubble radius increases (see figure).

\begin{figure}
\centerline{
\includegraphics[width=7.0cm]{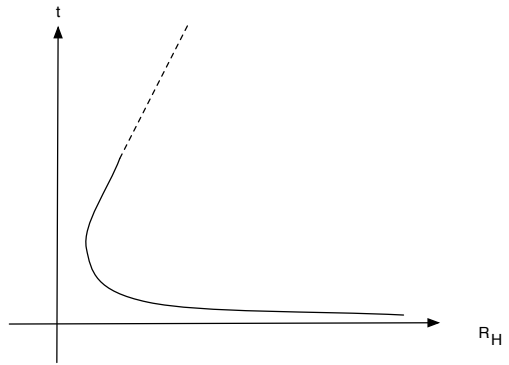}}
\caption{The time evolution of the Hubble radius in the tachyon cosmology with a spherical space.}
\end{figure}

Here, one should also check that in this period $\dot{B}^2$ given in \eq{tac1} is always  positive. To see this is indeed the case, let us define $f=\s e^{-2B}$ and $g=V/(6\sqrt{1-\dot{T}^2})$. Then, initially  $f(0)=g(0)$. From \eq{tac3} we find that
\be
\frac{d\ln g}{dt} = \frac{d\ln f}{dt} \,\left(\fr{3\dot{T}^2}{2}\right).\label{pos}
\ee
Since initially $\dot{B}(0)=0$ and $\ddot{B}(0)>0$, there is an interval during which $\dot{B}>0$ and consequently $d(\ln f)/dt<0$. Also from \eq{tac3}, we have
$d(\ln g)/dt<0$. Then \eq{pos} implies $d(\ln g)/dt>d(\ln f)/dt$ since initially $\dot{T}\ll 1$. Therefore, $g>f$ during this period and from \eq{tac1} one sees $\dot{B}^2>0$. Iterating, it is guaranteed that $\dot{B}^2>0$ until  $\dot{T}^2=2/3$.

As a result, we see that it is possible to obtain a cosmological phase with  decreasing Hubble radius which can be matched  to the static strongly coupled Hagedorn phase, provided that the flat space is replaced by a sphere, the tachyon field dominates the cosmic evolution and the initial tachyon potential energy is fine tuned with the curvature of the three sphere. Of course, it should be checked that these assumptions do not invalidate the results of \cite{pert1,pert2,pert3,pert4}. Especially, since the curvature of the sphere introduces a new length scale, it may alter the  scale independence of the perturbation spectrum. Moreover, the existence of the tachyon field should be explained. In any case, we find it interesting that a slight modification of the tachyon cosmology yields a decreasing Hubble radius.

\acknowledgments{The work of Ali Kaya is partially supported by Turkish Academy
  of Sciences via Young Investigator Award Program (T\"{U}BA-GEB\.{I}P).}

\end{document}